\documentclass[12pt,a4paper]{article}
\usepackage[english]{babel}
\usepackage[latin1]{inputenc}
\usepackage{graphicx}
\usepackage[T1]{fontenc}
\usepackage{amsmath}

\usepackage[paper=a4paper,tmargin=22mm,bmargin=15mm,lmargin=20mm,rmargin=20mm,headsep=25pt]{geometry}

\newcommand{\h}{\hspace{.5cm}}

\begin{document}
\date{}

\title{A discussion on approximations for the determination\\ of the maximum period of a simple pendulum}
\author{Rog\' erio Netto Suave\thanks{rnsuave@yahoo.com.br} \hspace{0.2cm  and \hspace{0.1cm} Jos\'e Alexandre Nogueira\thanks{jose.nogueira@ufes.br}}\\
[0.2cm]\small\emph{Departamento de F\'isica, Centro de Ci\^encias Exatas}\\
[0.1cm]\small\emph{Universidade Federal do Esp\'irito Santo -- UFES}\\
[0.1cm]\small\emph{29075-910 -- Vit\'oria -- ES -- Brasil}}

%%%%%%%%%%%%%%%%%%%%%%%%%%%%%%%%%%%%%%%%%%%%%%%%%%%%%%%%%%%%%%%%%%%%%

\maketitle

%%%%%%%%%%%%%%%%%%%%%%%%%%%%%%%%%%%%%%%%%%%%%%%%%%%%%%%%%%%%%%%%%%%%%%%%%%%%%%%%%%%%%%%%%%%%%%%%%%%%%%%%%%%%%%%%%%%%%%%%%%%%%%%%%%%%%%%%%%%%%%%%%%%%%%%%%%%%%%%%%%%%%%%%
\begin{abstract}
In this paper we deal with the care one must have in adopting approximations in regard with terms he chooses to leave behind in the particular case of the expression valid for the maximum period of a long pendulum oscillating near Earth's surface.
\end{abstract}
%%%%%%%%%%%%%%%%%%%%%%%%%%%%%%%%%%%%%%%%%%%%%%%%%%%%%%%%%%%%%%%%%%%%%%%%%%%%%%%%%%%%%%%%%%%%%%%%%%%%%%%%%%%%%%%%%%%%%%%%%%%%%%%%%%%%%%%%%%%%%%%%%%%%%%%%%%%%%%%%%%%%%%%%

%\newpage
%%%%%%%%%%%%%%%%%%%%%%%%%%%%%%%%%%%%%%%%%%%%%%%%%%

%%%%%%%%%%%%%%%%%%%%%%%%%%%%%%%%%%%%%%%%%%%%%%%%%%
\section{Introduction}
%%%%%%%%%%%%%%%%%%%%%%%%%%%%%%%%%%%%%%%%%%%%%%%%%%
\label{sec:Intro}

\h Almost ever all physical phenomena are treated in an approximate way. Sometimes effects of minor importance are neglected alongside the analysis made, other times higher order corrections are abandoned when expansions are developed. Thus, it is fundamental for undergraduate students of exact science courses, including Physics, to dominate the use of approximations, mainly to have the ability of deciding which terms are outside the validity domain of the approximations made and that could therefore be disregarded. Still, the student should know how to recognize if new analysed effects (particularly minor importance effects) have same order of magnitude of previously neglected ones and, consequently, they can not be neglected any more.

However, it is well-known to teachers the great difficulty that students have in dealing with approximations. To recognize terms that could be discarded can represent a considerable challenge for them. Mostly, their approach is done almost automatically without any concern with careful analysis so leading them to wrong results.

The determination of the maximum period of oscillation for a simple pendulum oscillating in the vicinity of Earth's surface is a good example of situations where watchfulness with treatment of approximations is absolutely necessary. At first, approximation for small oscillations is considered and terms of higher order in the oscillation angle $\theta$ are neglected. Subsequently, the limit of pendulum length tending to infinity is used to determine the maximum period of oscillation. But, as we will see later, this procedure must be cautiously conducted. This means that we have to take into account in what conditions the approximations are valid and to learn what is the physical meaning of such conditions.

As far as we know, Gough was who firstly in literature called attention that even for very small angles of oscillation the period of a simple pendulum in a perfect vacuum, suspended from a perfectly rigid support by an inextensible string is not $2 \pi \sqrt{\frac{L}{g}}$ due to the radius of curvature of the Earth~\cite{gough}.

The goal of this paper is to explicitly calculate the maximum period for a simple pendulum oscillating around the terrestrial surface and, trough it, discuss the heeds students must take in using approximations and the conditions in which neglected terms are outside the validity region of the approximation. Also, we intend to show how an ingenuous use of careless results can lead to mistaken conclusions.

Although the determination of the maximum period of a simple pendulum is not unpublished, it is poorly known and its calculation is not such a simple matter. Therefore an explicit demonstration showing the decomposition of the acting forces and the use of the Newton Second Law  is required. Moreover in the papers found in literature~\cite{gough,burko} is not discussed the validity of the approximations considered, as we intend to do in this work.

The idea of using the simple pendulum as scenario to discuss some specific subject is not new in the literature. In 2008, J. Clement used the simple pendulum in order to illustrate that students have misunderstood primitive concepts of Mechanics~\cite{clement}. Furthermore, the study of the simple pendulum is important, not only because is one of the most frequent problem found in the textbooks of undergraduate courses, but also because a lot of others problems can be reduced to a similar problem to that of the simple pendulum. Although apparently simple the study of a simple pendulum can be a very rich problem, as it can be seen in references [4-26].

The article is organized in the following way. In section \ref{period}  we calculate the period for amplitudes of small angles of a simple pendulum in Earth's surface vicinity taking account the effect of the finite curvature of the Earth.  In section \ref{maximum period} we derive the maximum period as the limit of the period of the simple pendulum when the length of its string goes to infinity. Finally, we present our final considerations and conclusions in section \ref{Concl}.

%%%%%%%%%%%%%%%%%%%%%%%%%%%%%%%%%%%%%%%%%%%%%%%%%%%%%%%%%%%%%%%%%%%%%%%%%%%%%%%%%%%%%%%%%%%%
\section{Period for a simple pendulum in Earth's surface vicinity}
\label{period}
%%%%%%%%%%%%%%%%%%%%%%%%%%%%%%%%%%%%%%%%%%%%%%%%%%%%%%%%%%%%%%%%%%%%%%%%%%%%%%%%%%%%%%%%%%%%

\h The physical case of interest here is a pendulum made of a negligible mass string of length $L$ and a point mass $m$ performing small oscillations nearby the terrestrial surface, that is, the simple pendulum.

For an observer at Earth's surface, the forces that act on $m$ are the gravitational attraction exerted by Earth, the tension provided by the string and the fictitious forces~\cite{30} associated to the frame of reference attached to the spinning Earth. These fictitious forces are the well-known Coriolis and centrifugal forces.

\begin{figure}[h!]
\centering
\includegraphics[scale=1]{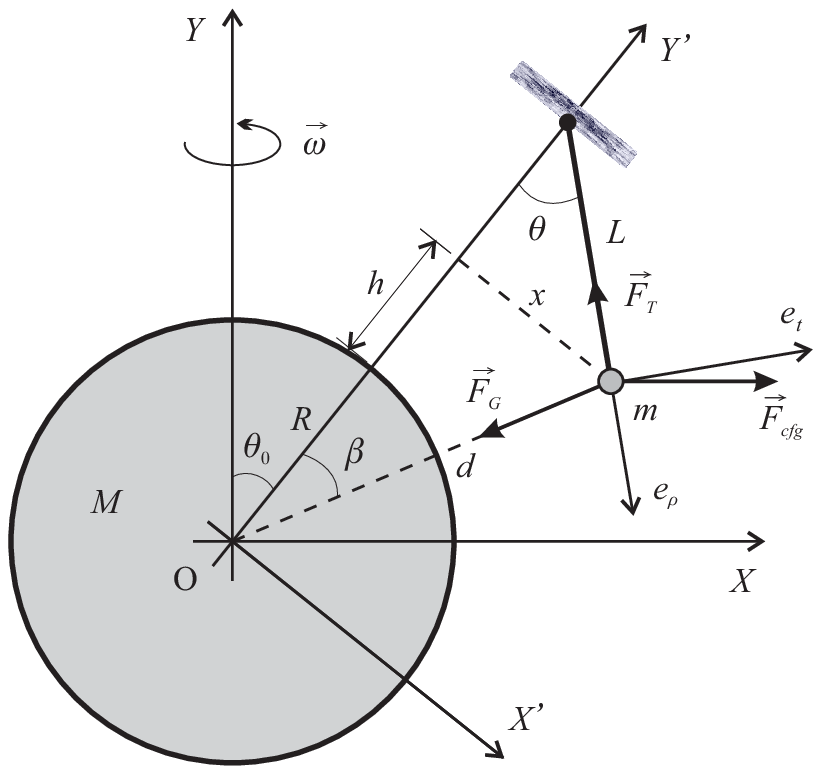}
\caption{Pendulum next to Earth's surface.}
\label{fig:ESpins}
\end{figure}

At Fig.~\ref{fig:ESpins} we represent the main axes necessary for a description of the problem. $Y$ is Earth's rotation axis and $X$ an axis orthogonal to $Y$ in such way that the pendulum -- mass and string -- is in plane $XY$ at the moment shown. Now, $Y'$ is an axis through Earth's centre and the fixation point of the string and $X'$ is orthogonal to $Y'$ also in the plane $XY$ at the instant shown. The figure indicates schematically the forces that act on $m$, except for the Coriolis force $\vec{F}_{Coriolis} = -2m\vec{\omega} \times \dot{\vec{r}^{\prime}}$ entering obliquely in the page, with perpendicular and parallel components relatively to the plane of the figure. The situation described here corresponds to the case where the trajectory's lowest point occur at latitude $\theta_{0}$\,, where the axis $e_{\rho}$ alongside the pendulum string turns into anti-parallel to $Y'$. The axis $e_{t}$ is perpendicular to $e_{\rho}$ and is also tangential to the trajectory of mass $m$.

%%%%%%%%%%%%%%%%%%%%%%%%%
%%%%%%%%%%%%%%%%%%%%%%%%%%%%

To avoid loosing the focus of this work, that is, to avoid deviating attention of the reader for the complexity that the solution becomes when fictitious forces are taken into account, rotation of Earth will be disregarded. The main advantage of neglecting action of Coriolis force is that the pendulum will oscillate at a fixed plane, so the mass path is a circle arc. That way, all axes in Fig.~\ref{fig:ESpins} will be at the same plane all the time, so one can define unit vectors $\hat{e}_{t}$ and $\hat{e}_{\rho}$ respectively tangent and perpendicular to the described arc at each instant of time. At last, with no loss of generality, one can consider also that this plane -- the oscillation plane -- is parallel to a terrestrial meridian (because of the spherical symmetry).

\begin{figure}[h!]
\centering
\includegraphics[scale=1]{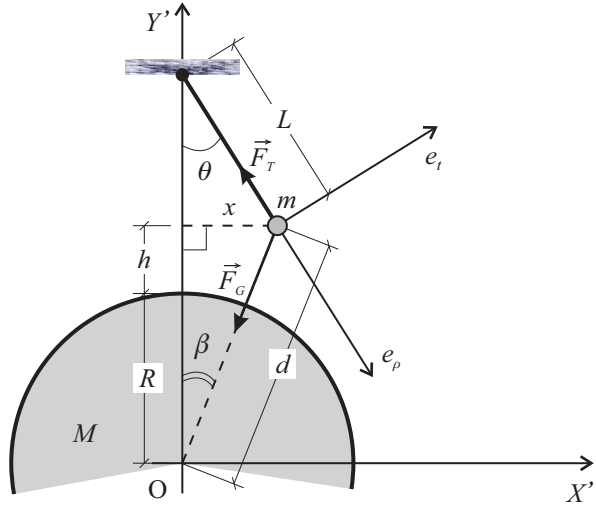}
\caption{Pendulum nearby the terrestrial surface disregarding Earth's rotation.}
\label{fig:EStops}
\end{figure}

With these simplifications, for an observer at the proximity of Earth's surface\cite{32} the forces that act on $m$ are the tension $\vec{F}_{T}$ along the string and the terrestrial gravitational attraction force $\vec{F}_{G}$ along local Earth's radius. These forces are shown at Fig.~\ref{fig:EStops} where we have rotated the coordinate system for representing axis $Y'$ at local vertical direction. This rotation was done for the sake of simplicity and for providing a better visualization.

%%%%%%%%%%%%%%%%%%%%%%%%%
%%%%%%%%%%%%%%%%%%%%%%%%%%%%

\begin{figure}[h!]
\centering
\includegraphics[scale=0.4]{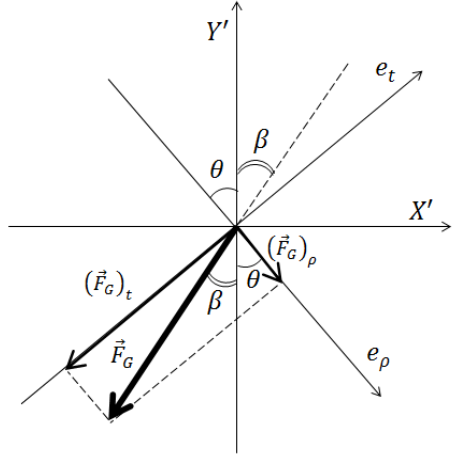}
\caption{Gravitational force components over axes $e_{t}$ and $e_{\rho}$.}
\label{fig:FgComp}
\end{figure}

For better understanding the analysis of pendulum oscillation that follows, it is convenient to decompose the gravitational force over the axes $e_{t}$ and $e_{\rho}$\,. Fig.~\ref{fig:FgComp} shows how the components alongside those axes are obtained (axes $X'$ e $Y'$ are also displayed for reference). Thus, as indicated at the figure, the force of gravitational attraction exerted by the Earth on mass $m$ is given by
\begin{equation}
\vec{F}_{G} = -{F}_{G} \sin{(\beta + \theta)} \:\hat{e}_{t} + {F}_{G} \cos{(\beta + \theta)} \:\hat{e}_{\rho} \:,
\end{equation}
that is,
\begin{equation}
\label{FGravt}
\vec{F}_{G} = -{F}_{G} \left( \cos{\beta}\sin{\theta} + \sin{\beta}\cos{\theta} \right)\hat{e}_{t} - {F}_{G} \left( \sin{\beta}\sin{\theta} - \cos{\beta}\cos{\theta} \right)\hat{e}_{\rho} \:.
\end{equation}
In this expression ${F}_{G}$ is the magnitude of the gravitational attraction given by Newton's Law of Universal Gravitation, so
\begin{equation}
\label{mgrv1}
{F}_{G} =G \frac{mM}{d^{2}} \:,
\end{equation}
where $G$ is the gravitational constant, $M$ is Earth's mass and $d$ is the distanced from $m$ to the centre of Earth (see Figs.~\ref{fig:ESpins} and \ref{fig:EStops}).

%%%%%%%%%%%%%%%%%%%%%%%%%
%%%%%%%%%%%%%%%%%%%%%%%%%%%%

The tension provided by the string on mass $m$ has the direction of axis $e_{\rho}$, therefore given by $\vec{F}_{T} = -F_{T} \:\hat{e}_{\rho}$\,.

Once we had established that $m$ oscillates nearby the terrestrial surface, as a first approximation we can neglect Earth's curvature and consider that $d \approx R$ and $\beta \approx 0$. Physically this is equivalent to consider the gravitational force always at the vertical direction $Y'$~\cite{33}. In this case, this force is nothing less than the weight of the point of mass $m$. With these considerations, Eq. (\ref{FGravt}) turns out to be
\begin{equation}
\vec{F}_{G} \approx -G \frac{mM}{R^{2}}\sin{\theta} \:\hat{e}_{t} + G \frac{mM}{R^{2}}\cos{\theta} \:\hat{e}_{\rho} \:.
\end{equation}

Applying Newton's Second Law to movement of mass $m$ for the tangential direction we have
\begin{equation}
L \:\frac{d^{2}\theta}{dt^{2}} \approx -g\sin{\theta} \:,
\end{equation}
where $g = G M / R^{2}$ is the magnitude of standard acceleration of gravity.

For small oscillations we may consider $\sin{\theta} \approx \theta$ and write
\begin{equation}
\label{EqMovNml}
\frac{d^{2}\theta}{dt^{2}} \approx -\frac{g}{L} \:\theta \:.
\end{equation}

Hence, the well-known result for the period of a simple pendulum undergoing small oscillations around Earth's surface is obtained as
\begin{equation}
\label{PeriodNml}
T_{0} \approx 2 \pi \sqrt{\frac{L}{g}} \:.
\end{equation}
Ergo, as a good approximation, this result shows that the period of a simple pendulum is independent on the amplitude of the small oscillations done.

A naive analysis of result (\ref{PeriodNml}) leads to the wrong conclusion that the period of a simple pendulum oscillating at vicinity of Earth's surface would increase indefinitely as the pendulum length $L$ increases. The reader is probably asking why wrong if by principle he has yet that $d \approx R$ and also that the angle $\beta$ still continues to be small with $L$ increasing, since the oscillations continue to occur at proximity of the terrestrial surface. As we will show at a later subsection, this is not always true because Earth's curvature has to be considered when the pendulum length is augmented excessively.

Figure~\ref{fig:EStops} depicts the details relating the angles of interest, from where we easily achieve
\begin{equation}
\label{BetaTheta}
\sin{\beta} =\frac{x}{d} = \frac{L}{d}\sin{\theta} \approx \frac{L}{R}\sin{\theta} \:.
\end{equation}
Therefore, when $L$ is not much lesser than $R$, the fraction $L/R$ is not much lesser than $1$ and $\sin{\beta}$ could not be negligible anymore. Furthermore, we remember that any value of $L$ can be used so $L/R$ can have any value, higher or much higher than $1$, so $\sin{\beta}$ could still be not inappreciable even in the case that $\theta$ is truly small.

%%%%%%%%%%%%%%%%%%%%%%%%%%%%%%%%%%%%%%%%%%%%%%%%%%%%%%%%%%%%%%%%%%%%%%%%%%%%%%%%%%%%
\subsection{Approximations for $\theta_{M}$ amplitudes of small angles}
%%%%%%%%%%%%%%%%%%%%%%%%%%%%%%%%%%%%%%%%%%%%%%%%%%%%%%%%%%%%%%%%%%%%%%%%%%%%%%%%%%%%

\h In this subsection, the term ``small oscillations'' is used to imply very small angles for the amplitude of oscillation, meaning that $\theta \leq \theta_{M}$ with $\theta_{M} \ll 1$\,.

\begin{figure}[h!]
\centering
\includegraphics[scale=1]{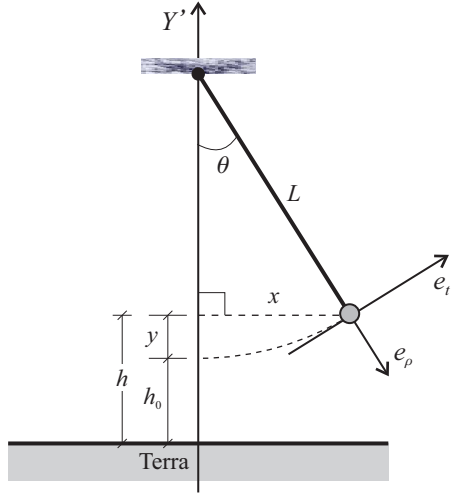} 
\caption{Relevant distances ($h$, $h_{0}$, and $y$) in connection to Earth's surface.}
\label{fig:GeoExp}
\end{figure}

According to Fig.~\ref{fig:EStops} the distance $d$ from the centre of Earth to point mass $m$ is related to the horizontal distance $x$ which separates $m$ from the vertical axis $Y'$ through

\begin{equation}
\label{dQd_ThetaPeq}
d^{2} = R^{2} \left( 1 + 2\frac{h}{R} + \frac{h^{2}}{R^{2}} \right)  + x^{2} \:,
\end{equation}
where $R$ is Earth's radius and $h$ is shown in Fig.~\ref{fig:GeoExp} (see also Fig.~\ref{fig:ESpins}).

%%%%%%%%%%%%%%%%%%%%%%%%%
%%%%%%%%%%%%%%%%%%%%%%%%%%%%

Figure~\ref{fig:GeoExp} has an expanded view of the geometry of oscillation above Earth's surface in which we introduce the vertical distance $y$ between the lower position that $m$ can have and its actual position at each instant of time. We then see that $y = L - L\cos{\theta}$ and $h = h_{0} + y$, where $h_{0}$ is the smallest distance between the terrestrial surface and mass $m$. In such way, we get
\begin{equation}
\label{h_ThetaPeq}
h = h_{0} + L\left( 1 - \cos{\theta} \right).
\end{equation}
Using Taylor's expansion of $\cos{\theta}$ in powers of $\theta$, equation above becomes\cite{36}
\begin{equation}
\label{hXp_ThetaPeq}
h = h_{0} + L \left[ \frac{\theta^{2}}{2} + {\cal O}\left( \theta^{4} \right) \right].
\end{equation}
For small oscillations, terms of higher order in $\theta$ are neglected so it is obtained
\begin{equation}
\label{hAp_ThetaPeq}
h \approx h_{0} + L \frac{\theta^{2}}{2} \:.
\end{equation}

From Fig.~\ref{fig:GeoExp} again it comes that
\begin{equation}
\label{x_HorzDist}
x = L \sin{\theta} \:.
\end{equation}
Now, expanding $\sin{\theta}$ in powers of $\theta$ leads to
\begin{equation}
\label{xXp_ThetaPeq}
x = L \left[ \theta +  {\cal O}\left( \theta^{3} \right) \right].
\end{equation}
Again, for small angle oscillations we throw away higher order terms in $\theta$ to obtain
\begin{equation}
\label{xAp_ThetaPeq}
x \approx L \theta \:.
\end{equation}

Substituting Eqs. (\ref{hAp_ThetaPeq}) and (\ref{xAp_ThetaPeq}) in expression (\ref{dQd_ThetaPeq}) we arrive at
\begin{equation}
d^{2} = R^{2} \left[ \left( 1 + 2\frac{h_{0}}{R} + \frac{h_{0}^{2}}{R^{2}} \right) + \left( 1  + \frac{h_{0}}{R} + \frac{L}{R} \right) \left( \frac{L}{R} \right) \theta^{2} + {\cal O}\left( \theta^{4} \right) \right].
\end{equation}

Since mass $m$ is in movement near Earth's surface it is valid to take $h_{0} \ll R$\,, that is $\frac{h_{0}}{R} \ll 1$\,. This implies that terms of ${\cal O}\left( \frac{h_{0}}{R} \right)$ are negligible when compared to $1$ in first and second sets of parenthesis of the above expression. Being so, up to terms of order $\theta^2$ we have
\begin{equation}
\label{dXp_ThetaPeq}
d = R \left[ {1 + \left( \frac{L}{R}  + \frac{L^{2}}{R^{2}} \right) \theta^{2}} \right]^{1/2} ,
\end{equation}
and
\begin{equation}
\label{dXp_ThetaPeq_inv}
\frac{1}{d} = \frac{1}{R} \left[ 1 - \left( \frac{L}{R} + \frac{L^{2}}{R^{2}} \right) \frac{\theta^{2}}{2} \right].
\end{equation}
In principle, terms of ${\cal O}\left( \theta^{2} \right)$ can be neglected in Eq. (\ref{dXp_ThetaPeq}) (and (\ref{dXp_ThetaPeq_inv})) for small oscillations of the pendulum. However, we will show later that this is an ingenuous procedure and such approach must be carefully handled.

Otherwise, we follow with that naive supposition about the validity domain of the approximations and we get
\begin{equation}
\label{dAp_ThetaPeq}
d = R \:.
\end{equation}

Now, from Eqs. (\ref{xXp_ThetaPeq}) and (\ref{dXp_ThetaPeq_inv}) we obtain for $\sin{\beta}$ that
\begin{equation}
\label{SinBeta_ThetaPeq}
\sin{\beta} = \frac{x}{d} \approx \frac{L}{R} \left[ 1 - \left( \frac{L}{R} + \frac{L^{2}}{R^{2}} \right)\frac{\theta^{2}}{2} \right]\theta.
\end{equation}
After throwing away terms of ${\cal O}\left( \theta^{3} \right)$ (or using approximations (\ref{xAp_ThetaPeq}) and (\ref{dAp_ThetaPeq})) we obtain
\begin{equation}
\label{SinBetaAp_ThetaPeq}
\sin{\beta} \approx \frac{L}{R} \theta \:.
\end{equation}
Also, an approximation up to second order in $\theta$ for $\cos{\beta}$ comes from
\begin{equation}
\cos{\beta} = \left( 1 - \sin^{2}{\beta} \right)^{1/2} \approx \left( 1 -\frac{L^{2}\theta^{2}}{R^{2}} \right)^{1/2} ,
\end{equation}
that is,
\begin{equation}
\label{CosBetaAp_ThetaPeq}
\cos{\beta} \approx 1 - \frac{L^{2}}{2R^{2}}\theta^{2} \:.
\end{equation}
Thus, for the approximation considered so far we throw away terms of ${\cal O}\left( \theta^{2} \right)$ and continue using $\cos{\beta} = 1$.

This way, the tangential component of the gravitational force (\ref{FGravt}) becomes
\begin{equation}
\label{FGravt_t_ThetaPeq}
(\vec{F}_{G})_{t} \approx -mg \left( 1 + \frac{L}{R} \right)\theta \:\hat{e}_{t} \:.
\end{equation}

Considering only tangential motion, we apply Newton's Second Law on mass $m$ to get
\begin{equation}
\label{EqMovMod}
\frac{d^{2}\theta}{dt^{2}} \approx -\frac{g}{L} \left( 1 + \frac{L}{R} \right)\theta \:.
\end{equation}

Equation (\ref{EqMovMod}) shows that, in the order which we are considering, the mass $m$ undergoes a simple harmonic motion in respect to $\theta$ (or in respect to arc $s =L \, \theta$) with a period of oscillation given by
\begin{equation}
\label{PeriodMod}
T = 2\pi \sqrt{\frac{L}{g\left( 1 + L/R \right)}} \:.
\end{equation}

%%%%%%%%%%%%%%%%%%%%%%%%%
\begin{figure}[h!]
\centering
\includegraphics[scale=1]{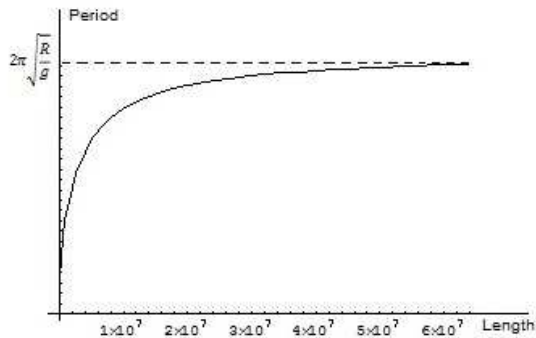} 
\caption{Corrected period for a long pendulum, showing asymptotic behaviour to the limit $T_{max} = 2 \pi \sqrt{R/g}$ when $L \rightarrow \infty$\,.}
\label{fig:grafico1}
\end{figure}
%%%%%%%%%%%%%%%%%%%%%%%%%%%%

It is convenient to change to a dimensionless length $u := L/R$ and to rewrite the oscillation period as
\begin{equation}
\label{PeriodMod_Adim}
T = 2\pi \sqrt{\frac{R}{g}} \left( \frac{u}{u+1} \right) ^{1/2} \:.
\end{equation}

%%%%%%%%%%%%%%%%%%%%%%%%%%%%%%%%%%%%%%%%%%%%%%%%%%%%%%%%%%%%%%%%%%%%%%%%%%%%%%%%%%%%%%%%%%%%
\section{Maximum Period}
\label{maximum period}
%%%%%%%%%%%%%%%%%%%%%%%%%%%%%%%%%%%%%%%%%%%%%%%%%%%%%%%%%%%%%%%%%%%%%%%%%%%%%%%%%%%%%%%%%%%%

\h Then, unlikely to the prediction of Eq. (\ref{PeriodNml}) that the period of the pendulum oscillating near Earth's surface would increase unlimitedly with pendulum length, Eqs. (\ref{PeriodMod}) or (\ref{PeriodMod_Adim}) above show that in fact the period approaches asymptomatically to a maximum value $T_{max}$ established by the limit where $L \rightarrow \infty$ or $u \rightarrow \infty$,
\begin{equation}
\label{tmax}
T_{max} = 2\pi \sqrt{\frac{R}{g}}.
\end{equation}

To get an estimate, we use the mean Earth's radius $\bar{R}=6.371\times10^{6}\,\text{m}$ and the standard value for the acceleration of gravity $g=9.807\,\text{ms}^{-2}$ to achieve $T_{max} \approx 5,064\,\text{s} \approx 84.40\,\text{min}$~\cite{romer}.

%%%%%%%%%%%%%%%%%%%%%%%%%
\begin{figure}[h!]
\centering
\includegraphics[scale=1]{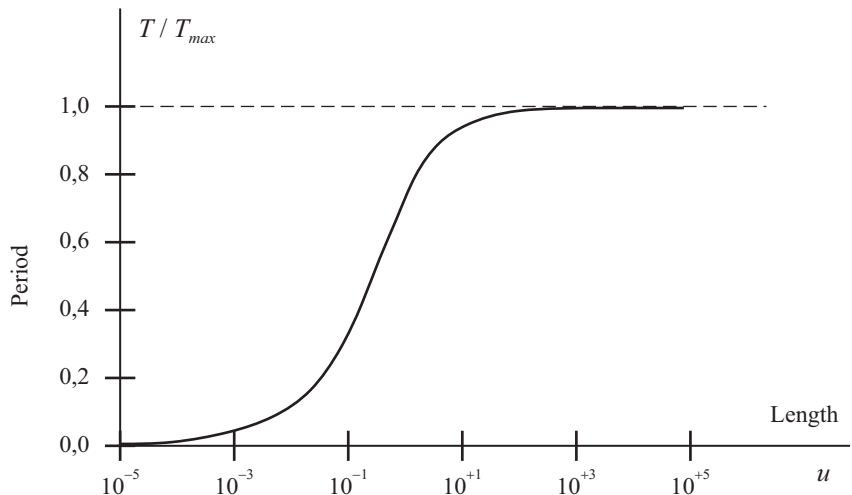} 
\caption{Corrected period for a long pendulum, showing asymptotic behaviour to the limit $T_{max} = 2 \pi \sqrt{R/g}$ when $u \rightarrow \infty$\,.}
\label{fig:PerVsL}
\end{figure}
%%%%%%%%%%%%%%%%%%%%%%%%%%%%

Figures \ref{fig:grafico1} and \ref{fig:PerVsL} show this behaviour. In Fig. \ref{fig:PerVsL}, axis $u$ is in a logarithmic scale to better exhibit the asymptotic trend as $u \rightarrow \infty$\,. At the limits $u \ll 1$, the expression above (\ref{PeriodMod_Adim}) leads to $T \approx T_{0}$ showing that it predicts the adequate behaviour.
%The dashed line in the Fig. shows the inadequate prediction of previous expression (\ref{PeriodNml_Adim}) that $T_{0} \rightarrow \infty$ when $u \rightarrow \infty$\,.

Although seemingly reasonable, result (\ref{tmax}) was obtained in an ingenuous analysis and could be not valid even for small angular amplitudes of oscillation.  The mistake made is in the fact that terms of ${\cal O}\left( \theta^{2} \right)$ (and of higher orders) were neglected in Eqs. (\ref{dXp_ThetaPeq_inv}), (\ref{SinBeta_ThetaPeq}) and (\ref{CosBetaAp_ThetaPeq}) and in the fact that the coefficients of those terms depend on the parameter $L/R$ which, as already depicted, can assume any value. This occurs because the fraction $\frac{L}{R}$ increases with increasing $L$ so the first terms thrown away of $d$, $\sin{\beta}$ and $\cos{\beta}$ could become of order $\theta$ or of order smaller than $\theta$, facts that would make the approximations invalid. It is easy to notice that if $L$ increases up to $\frac{L}{R} \approx \frac{1}{\theta_{M}}$ is satisfied, the term $\frac{L^{2}}{R^{2}}\theta^{2}$ is no longer negligible in comparison with $1$. Consequently, this term is no longer outside the validity domain of the approximation. This way, a more cautious approach is needed and one must impose that the approximation will be applicable only if it is satisfied the condition
\begin{equation}
\label{CondConvg}
\frac{L}{R} \ll \frac{1}{\theta_{M}} \:,
\end{equation}
where $\theta_{M}$ is the angular amplitude of pendulum oscillation. The reader may be questioning if there are not terms $\frac{L}{R}$ with higher powers that could dominate for large values of $L$. But, the reader can verify that the substitution of expressions (\ref{hXp_ThetaPeq}) and (\ref{xXp_ThetaPeq}) in Eq. (\ref{dQd_ThetaPeq}) conducts to
\begin{equation}
d = R \left\{ 1 +  \frac{L}{R} \left[ \theta^{2} +  {\cal O}\left( \theta^{4} \right) \right] + \frac{L^{2}}{R^{2}} \left[ \theta^{2} + {\cal O}\left( \theta^{4} \right) \right] \right\}^{1/2} .
\end{equation}
Then, for small angles $\theta$, meaning $|\theta| \leq \theta_{M}$ with $\theta_{M} \ll 1$, it is possible to throw away terms of ${\cal O}\left( \theta^{4} \right)$ in comparison to those of order $\theta^{2}$, which assures the validity of the condition (\ref{CondConvg}). Moreover, the significant terms in the remaining expansions are of form $\left( \frac{L \theta^{n}}{R} \right)^{2}$ which stay lesser than $1$ for values of $L$ satisfying condition (\ref{CondConvg}).

\begin{figure}[h!]
\centering
\includegraphics[scale=1]{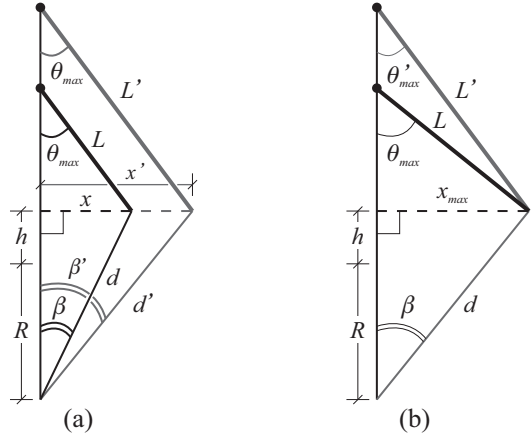} 
\caption{Two situations showing the effect of varying the pendulum length. In (a) the length is varied against keeping fixed the angle of oscillation $\theta_{M}$. Otherwise, in (b) it is the linear amplitude $x_{max}$ which is kept fixed.}
\label{fig:GeoVsL}
\end{figure}

We can do some physical analysis to better understand what happens. If the pendulum length is augmented keeping angular amplitude fixed, even for small $\theta_{M}$ the distance $x_{max}$ (horizontal displacement amplitude) is allowed to increase too much in a way that mass $m$ departs too much from the surface in view of Earth's curvature. In such case, the distance $d$ from Earth's centre to $m$ is no longer close to $R$, nor the angle $\beta$ can be considered small. Figure ~\ref{fig:GeoVsL}(a) is a way of displaying the situation here described, by showing two representations of relevant distance/angle magnitudes with the same angle of oscillation $\theta_{M}$ but with different lengths of the pendulum.

The physical analysis made in the previous paragraph outlines us a tip of how to make a feasible approach to avoid the troubles we have mentioned there. If the amplitude $x_{max}$ of horizontal displacement of $m$, as measured from axis $Y'$, is kept fixed then the angular amplitude $\theta_{M}$ must be reduced when pendulum length $L$ is increased, and this way the condition (\ref{CondConvg}) is assured. This situation is made explicit at Fig.~\ref{fig:GeoVsL}(b), showing that in order to maintain $x_{max}$ constant $\theta_{M}$ diminishes as $L$ increases.

%%%%%%%%%%%%%%%%%%%%%%%%%
%%%%%%%%%%%%%%%%%%%%%%%%%%%%

%%%%%%%%%%%%%%%%%%%%%%%%%%%%%%%%%%%%%%%%%%%%
\section{Final considerations}
\label{Concl}
%%%%%%%%%%%%%%%%%%%%%%%%%%%%%%%%%%%%%%%%%%%%

\h The main goal of this work is to discuss the use of approximations in solving physical problems, calling attention mainly to the care one must have when throwing away terms, namely, how to choose among the terms derived which ones can be considered negligible in order to get a good usage and a good validity of the approximate results obtained.

Even though the problem so far studied can be thought of as an academic one it is by no means a poor example. In fact, the determination of maximum period for a simple pendulum oscillating nearness the terrestrial surface illustrates in a simple way the essential points which we intend to call close attention from the reader. In determining the pendulum period, terms of ${\cal O} \left(\theta^{2} \right)$ are neglected when the pendulum undergoes small oscillations. However, we have shown that this ought to be done with some precaution since several seemingly terms are in fact of ${\cal O} \left(\frac{L}{R} \theta^{2} \right)$ or ${\cal O} \left(\frac{L^{2} \theta^{2}}{R^{2}} \right)$. Therefore, while taking the limit $L \rightarrow \infty$ to get the maximum period we are really admitting that $L$ can increase with no bounds and, in such circumstance, the ratio $\frac{L}{R}$ increases so much that those cited terms can become no more inappreciable even for the case of small oscillations. Put this way, these terms could have the same order of magnitude than terms regularly kept in approximate expressions, that is, in the normal range of the validity region. Thus, to assure validity of Eq. (\ref{PeriodMod}) and also of the intended limit to $T_{max}$ it is necessary to impose that $\frac{L}{R} \ll \frac{1}{\theta_{M}}$. We have shown that this is equivalent to make physical angular amplitudes $\theta_{M}$ decrease with increase of $L$. From this condition we proclaim that, in order to obtain the maximum period of the simple pendulum at vicinity of Earth's surface, the term ``small oscillations'' is to be understood as fixed small amplitudes of the horizontal displacement $x_{max}$. Therefore, in this case, it is effectively valid to take the limit $L \rightarrow \infty$ to find the maximum period of oscillation $T_{max} = 2 \pi (R/g)^{1/2}$ since even if the length $L$ increases indefinitely the result (\ref{PeriodMod}) remains inside the domain of validity for the approximations so far considered.

As we said at beginning of section \ref{sec:Intro} not only terms of higher order but also those of lesser importance were abandoned. So, we think it is worthwhile do discuss briefly how the effects of terrestrial rotation can influence the results gotten so far. The centrifugal force that acts on mass $m$ modifies the period with a correction term of smaller order which values at most $\frac{R \omega^{2}}{g} \approx 10^{-4}$ (at the equatorial line). Thus, this will be a dominant term for small lengths and the behaviour shown at Figs.~\ref{fig:PerVsL} or \ref{fig:grafico1} for small $L$ is not reliable. Nevertheless, the term $\frac{L}{R}$ dominates when the length increases too much. This way, we roughly conclude that effects of the centrifugal acceleration over the determination of the pendulum period are out of the region of validity for the approximations and the result (\ref{tmax}) is fully adequate (if condition $\frac{L}{R} \theta \ll 1$ is assured). Besides, as is well-known~\cite{marion,fowles, greiner}, the Coriolis force causes mass $m$ to not oscillate in a fixed plane so the ulterior motion can be described as a combination of two independent simple harmonic motions with equal periods $T = 2 \pi (L/g)^{1/2}$. This is valid for the Coriolis contribution when small oscillations are analysed and when Earth's curvature is not considered. Curvature effects upon the period of oscillation are too complex to be dealt with and are out of this work's aims.

Finally, if the effect of finite curvature of the Earth is sufficiently relevant to be taken into account then the effect of the gravitational gradient must also be taken into account in determination of the period of the simple pendulum, that is, the effect on $g$ due to the non-homogeneity of the Earth \cite{suits}.

%%%%%%%%%%%%%%%%%%%%%%%%%%%%%%%%%%%%%%%%%%%%%%%%%%%%%%%%%%%%%%%%%%%%%%%%%%%%%%%%%%%%%%%%%%%%%%%%%%%%%%%%%%%%%%%%%%%%%%%%%%%%%%%%%%%%%%%%%%%%%%%%%%%%%%%%%%%%%%%%%%%%%%%%%
%%%%%%%%%%%%%%%%%%%%%%%%%%%%%%%%%%%%%%%%%%%%%%%%%%%%%%%%%%%%%%%%%%%%%%%%%%%%%%%%%%%%%%%%%%%%%%%%%%%%%%%%%%%%%%%%%%%%%%%%%%%%%%%%%%%%%%%%%%%%%%%%%%%%%%%%%%%%%%%%%%%%%%%%%

%%%%%%%%%%%%%%%%%%%%%%%%%%%%%%%%%%%%%%%%%


\begin{thebibliography}{99}

\bibitem{gough} W. Gough, ``The period of a simple pendulum is not $2 \pi \sqrt{\frac{l}{g}}$,'' Eur. J. Phys. {\bf 4}, 53-54 (Jan.1983).

\bibitem{burko} L. M. Burko, ``Effect of the spherical Earth on a simple pendulum,'' Eur. J. Phys. {\bf 24}, 125-130 (Jan.2003).

\bibitem{clement} J. Clement, ``Students' preconceptions in introductory mechanics,'' Am. J. Phys. {\bf 50}, 66-71 (Jan.1982).

\bibitem{fulcher} L. P. Fulcher and B. F. Davis, ``Theoretical and experimental study of the motion of the simple pendulum,'' Am. J. Phys. {\bf 44}, 51-55 (Jan.1976).

\bibitem{suppes} C. G. Carvalhaes and P. Suppes, ``Approximations for the period of the simple pendulum based on the arithmetic-geometric mean,'' Am. J. Phys. {\bf 76}, 1150-1154 (Dec.1976).

\bibitem{arons} A. B. Arons, ``How long is a simple pendulum?,'' Phys. Teach. {\bf 15}, 300-301 (May.1977).

\bibitem{ganley} W. P. Ganley, ``Simple pendulum approximation,'' Am. J. Phys. {\bf 53}, 73-76 (Jan.1985).

\bibitem{nelsonolsson} R. A. Nelson and M. G. Olsson, ``The pendulum-Rich physics form a simple system,'' Am. J. Phys. {\bf 54}, 112-121 (Feb.1986).

\bibitem{cadwelboyco} L. H. Cadwell and E. R. Boyco, ``Linearization of the simple pendulum,'' Am. J. Phys. {\bf 59}, 979-981 (Nov.1991).

\bibitem{cromer} A. Cromer, ``Many oscillations of a rigid rod,'' Am. J. Phys. {\bf 63}, 112-121 (Feb.1995).

\bibitem{molina} M. I. Molina, ``Simple Linearizations of the Simple Pendulum for any Amplitude,'' Phys. Teach. {\bf 35}, 489-490 (Nov.1997).

\bibitem{randall} Randall D. Peters, ``Student-Friendly Precision Pendulum,'' Phys. Teach. {\bf 33}, 390-393 (Oct.1999).

\bibitem{moreland} P. Moreland, ``Improving Precision and Accuracy in the$g$ Lab,'' Phys. Teach. {\bf 38}, 367-369 (Sept.2000).

\bibitem{erkat} C. Erkat, ``The simple pendulum: a relativistic revisit,'' Eur. J. Phys. {\bf 21}, 377-384 (Sept.2000).

\bibitem{kiddfogg} R. B. Kidd and S. L. Fogg, ``A Simple Formula for the Large-Angle Simple Pendulum,'' Phys. Teach. {\bf 40}, 81-83 (Feb.2002).

\bibitem{millet} L. E. Millet, ``The Large-Angle Pendulum Period,'' Phys. Teach. {\bf 41}, 162-163 (Mar.2003).

\bibitem{parwani} R. R. Parwani, ``An approximation expression for the large angle of a simple pendulum,'' Eur. J. Phys. {\bf 25}, 37-39 (Oct.2004).

\bibitem{aggarwal} N. Aggarwal, N. Verma and P. Arun, ``Simple pendulum revisited,'' Eur. J. Phys. {\bf 26}, 517-523 (Apr.2005).

\bibitem{hite} G. E. Hite, ``Approximations for the period of a simple pendulum,'' Phys. Teach. {\bf 43}, 290-292 (May.2005).

\bibitem{belendez} A. Bel\'endez, A. Hern\'andez, A. M\'arquez, T. Bel\'endez and C. Neipp, ``Analytical approximations for the period of a nonlinear pendulum,'' Eur. J. Phys. {\bf 27}, 539-551 (Mar.2006).

\bibitem{limaarun} F. M. S. Lima and P. Arun, ``An accurate formula for the period of a simple pendulum oscillating beyond the small angle regime,'' 
Am. J. Phys. {\bf 74}, 892-895 (Oct.2006).

\bibitem{amore} P. Amore, M. C. Valdovinos, G. Ornelas, and S. Z. Barajas, ``The nonlinear pendulum: formulas for the large amplitude period,'' Rev. Mex. Fis E {\bf 53}, 106-111 (Jun.2007).

\bibitem{amrani} D. Amrani, P. Paradis and M. Beaudin, ``Approximation expression for the large-angle period of a simple pendulum revisited,'' Rev. Mex. Fis E {\bf 54}, 59-64 (Jun.2008).

\bibitem{belendez2} A. Bel\'endez, J. J. Rodes, T. Bel\'endez and A. Hern\'andez, ``Approximations for a large-angle simple pendulum period,'' Eur. J. Phys. {\bf 30}, L25-L28 (Feb.2009).

\bibitem{belendez1} A. Bel\'endez, J. Fran\'es, M. Ortu\~no, S. Gallego and J. G. Bernabeu, ``Higher accurate approximate solutions for the simple pendulum in terms of elementary functions,'' Eur. J. Phys. {\bf 31}, L65-L70 (Apr.2010).

\bibitem{Turkyilmazoglu} M. Turkyilmazoglu, ``Improvements in the approximate formulae for the period of the simple pendulum,'' Eur. J. Phys. {\bf 31}, 1007-1011 (Jul.2010).

\bibitem{30} Due to Earth's rotation, a reference system fixed on Earth's surface is non-inertial. The movement is conveniently analysed as a combination of a rotation with angular velocity $\vec{\omega}$, considering the origins of referentials coincident, and a translation of its axes - a uniform circular motion - with velocity $\vec{\omega} \times \vec{r}$.

\bibitem{32} At an inertial reference frame, since Earth's motion is disregarded.

\bibitem{33} Unless for irregularities at ground the effective acceleration of gravity is always perpendicular to the terrestrial surface even if we consider that Earth is rotating.

\bibitem{36} ${\cal O}(\theta^4)$ represents terms of order 4 or higher, i. e., terms of power $\theta^n$ where $n$ is an integer and $n \geq 4$.

\bibitem{romer} Robert H. Romer, ``The Answer Is Forty-Two -- Many Mechanics Problems, Only One Answer,'' Phys. Teach. {\bf 41}, 286-290 (May.2003).

\bibitem{marion} S. T. Thornton and J. B. Marion, {\it Classical Dynamics of Particles and Systems}, 5th ed. (Thomsom Books, New York, 2003).

\bibitem{fowles} G. R. Fowles and G. L. Cassiday, {\it Analytical Mechanics}, 7th ed. (Saunders College Publishing, New York, 2004).

\bibitem{greiner} Walter Greiner, {\it CLASSICAL MECHANICS -- Systems of Particles and Hamiltonian Dynamics}, 2nd ed (Springer-Verlag, New York, 2010).

\bibitem{suits} B. H. Suits, ``Long pendulums in gravitational gradients,'' Eur. J. Phys. {\bf 27}, L7-L11 (Jan.2006).




\end{thebibliography}
\end{document}